\documentclass{bmcart}

\usepackage[utf8]{inputenc} %unicode support
\usepackage{graphicx}

\startlocaldefs
\endlocaldefs

\begin{document}

%%% Start of article front matter
\begin{frontmatter}

\begin{fmbox}
\dochead{Research}


\title{The distorted mirror of Wikipedia: a quantitative analysis of Wikipedia coverage of academics}

%%%%%%%%%%%%%%%%%%%%%%%%%%%%%%%%%%%%%%%%%%%%%%
%%                                          %%
%% Enter the authors here                   %%
%%                                          %%
%% Specify information, if available,       %%
%% in the form:                             %%
%%   <key>={<id1>,<id2>}                    %%
%%   <key>=                                 %%
%% Comment or delete the keys which are     %%
%% not used. Repeat \author command as much %%
%% as required.                             %%
%%                                          %%
%%%%%%%%%%%%%%%%%%%%%%%%%%%%%%%%%%%%%%%%%%%%%%
  
\author[
   %addressref={aff1},                   % id's of addresses, e.g. {aff1,aff2}
   corref={aff1},                       % id of corresponding address, if any
   %noteref={n1},                        % id's of article notes, if any
   %email={ann.samoilenko@gmail.com}   % email address
]{\inits{AS}\fnm{Anna} \snm{Samoilenko}}
\author[
   addressref={aff1},
   email={taha.yasseri@oii.ox.ac.uk}
]{\inits{TY}\fnm{Taha} \snm{Yasseri}}

%%%%%%%%%%%%%%%%%%%%%%%%%%%%%%%%%%%%%%%%%%%%%%
%%                                          %%
%% Enter the authors' addresses here        %%
%%                                          %%
%% Repeat \address commands as much as      %%
%% required.                                %%
%%                                          %%
%%%%%%%%%%%%%%%%%%%%%%%%%%%%%%%%%%%%%%%%%%%%%%

\address[id=aff1]{%                           % unique id
  \orgname{Oxford Internet Institute, University of Oxford}, % university, etc
  \street{1 St Giles'},                     %
  \postcode{OX1 3JS}                                % post or zip code
  \city{Oxford},                              % city
  \cny{UK}                                    % country
}

%%%%%%%%%%%%%%%%%%%%%%%%%%%%%%%%%%%%%%%%%%%%%%
%%                                          %%
%% Enter short notes here                   %%
%%                                          %%
%% Short notes will be after addresses      %%
%% on first page.                           %%
%%                                          %%
%%%%%%%%%%%%%%%%%%%%%%%%%%%%%%%%%%%%%%%%%%%%%%

%\note{Sample of title note}     % note to the article
%\note[id=n1]{Equal contributor} % note, connected to author

\end{fmbox}% comment this for two column layout

%%%%%%%%%%%%%%%%%%%%%%%%%%%%%%%%%%%%%%%%%%%%%%
%%                                          %%
%% The Abstract begins here                 %%
%%                                          %%
%% Please refer to the Instructions for     %%
%% authors on http://www.biomedcentral.com  %%
%% and include the section headings         %%
%% accordingly for your article type.       %%
%%                                          %%
%%%%%%%%%%%%%%%%%%%%%%%%%%%%%%%%%%%%%%%%%%%%%%

\begin{abstractbox}

\begin{abstract} % abstract   max 300 w
Activity of modern scholarship creates online footprints galore. Along with traditional metrics of 
research quality, such as citation counts, online images of researchers and institutions increasingly 
matter in evaluating academic impact, decisions about grant allocation, and promotion. We examined 400 
biographical Wikipedia articles on academics from four scientific fields to test if being featured in 
the world's largest online encyclopedia is correlated with higher academic notability (assessed through 
citation counts). We found no statistically significant correlation between Wikipedia articles metrics 
(length, number of edits, number of incoming links from other articles, etc.) and academic notability of the mentioned researchers. 
We also did not find any evidence that the scientists with better WP representation are necessarily more prominent in their fields. 
In addition, we inspected the Wikipedia coverage of notable scientists sampled from Thomson Reuters
list of ``highly cited researchers''.
In each of the examined fields, 
Wikipedia failed in covering notable scholars properly. Both findings imply that Wikipedia might be producing an inaccurate image of academics on the front 
end of science. By shedding light on how public perception of academic progress is formed, this study alerts that a subjective element might have been introduced
into the hitherto structured system of academic evaluation.

\end{abstract}

%%%%%%%%%%%%%%%%%%%%%%%%%%%%%%%%%%%%%%%%%%%%%%
%%                                          %%
%% The keywords begin here                  %%
%%                                          %%
%% Put each keyword in separate \kwd{}.     %%
%%                                          %%
%%%%%%%%%%%%%%%%%%%%%%%%%%%%%%%%%%%%%%%%%%%%%%


\end{abstractbox}
%
%\end{fmbox}% uncomment this for twcolumn layout

\end{frontmatter}

%%%%%%%%%%%%%%%%%%%%%%%%%%%%%%%%%%%%%%%%%%%%%%
%%                                          %%
%% The Main Body begins here                %%
%%                                          %%
%% Please refer to the instructions for     %%
%% authors on:                              %%
%% http://www.biomedcentral.com/info/authors%%
%% and include the section headings         %%
%% accordingly for your article type.       %%
%%                                          %%
%% See the Results and Discussion section   %%
%% for details on how to create sub-sections%%
%%                                          %%
%% use \cite{...} to cite references        %%
%%  \cite{koon} and                         %%
%%  \cite{oreg,khar,zvai,xjon,schn,pond}    %%
%%  \nocite{smith,marg,hunn,advi,koha,mouse}%%
%%                                          %%
%%%%%%%%%%%%%%%%%%%%%%%%%%%%%%%%%%%%%%%%%%%%%%

%%%%%%%%%%%%%%%%%%%%%%%%% start of article main body
% <put your article body there>

%%%%%%%%%%%%%%%%
%% Background %%
%%
\section*{Introduction}
Modern scholarship is undergoing a revolutionary process of transformation triggered by the
advances in information and communication technology. In growing numbers, scholars are
moving their everyday work to the Web, creating diverse digital footprints galore. Recent
studies show that social media have become indispensable in supporting research related
activities \cite{1,2}. According to an analysis of STI conference presenters, 84\% of scholars have web
pages, 70\% are on LinkedIn, 23\% have public Google Scholar profiles, and 16\% are on Twitter.
Online reputation management is becoming essential in academic circles\cite{3}: 77\% of researchers
monitor their personal online images, and 88\% guard the reputation of their work online \cite{4}.
Researchers are advised to establish a Web presence on social media websites such as Twitter and 
Google+ so that they appear higher in the search results and thereby become more
visible \cite{6}.

These developments in the research enterprise, affect both formal (among scholars) and
informal (with the wider public) scholarly communication \cite{7,8}, and create new possibilities as
well as challenges, in the evaluation of the contribution of the individual researchers and the
scholarly progress in general \cite{9,10}. Modern research communities are under increasing pressure
to justify their scientific and societal value to the general public, funding agencies, and other
stakeholders, and online presence plays an important role in this competitive race \cite{10}.
These days citation analysis, which involves counting how many times a paper or a researcher is
referred to by other researchers, along with analyzing authors' citation networks, is
increasingly used to quantify the importance of scientists across the disciplines \cite{11}. Direct citation
counts and their functions like $h$-index \cite{12} and $g$-index \cite{13} are widely employed for scientific impact
evaluation and measuring researchers' visibility \cite{14}. They can be of fundamental importance in
decisions about hiring \cite{15} and grant awards \cite{16}, and often are the only way for non-specialists from
different fields or non-academic institutions to judge the impact of a scientific publication or a
scholar \cite{14}.

Although citation counts are universally acknowledged as the indicator of academic prestige,
they are loosely correlated with the future scientific impact of scientists \cite{17}. Previous research
admits their biases associated with (1) negative or ceremonial citations \cite{14}, (2) geographical
gravity laws in citation practices \cite{18}, (3) incomparability of citation counts across fields \cite{11} and
bibliographical databases \cite{14,20,21,22}. 
Finally, subject to delays caused by publishing and peer
review procedures, citation counts accumulate slowly and lag behind by several years.
Along with citation counts, public engagement is another factor essential for scientists' future
funding, promotions, and academic visibility \cite{6,23}. Increasingly often, this engagement is
happening online through different social media, and can be measured with the number of
``likes'', clicks, comments, downloads, ``retweets'', etc., which have been labeled {\it altmetrics}
(alternative metrics) \cite{24}. There is a wide scope of ongoing research exploring the role of social
media and altmetrics in providing alternatives to the traditional research evaluation \cite{3,25,26},
primarily, exploring how much altmetrics data exist \cite{27} and whether they can be used in
evaluating academic impact \cite{28}. For example, a recent study has found that earlier article
\cite{25}
download metrics can be used for predicting the future impact of academic papers, and some
researchers even suggest that academics should include altmetrics in their CVs \cite{23} as an
innovative indicator of academic importance.

In this study, our attention was drawn by Wikipedia, a web-based encyclopedia which allows
any user to freely edit its content, create and discuss the articles – all in the absence of central
authority or stable membership. This model of a decentralised bottom-up knowledge
construction draws on the wisdom of the crowds rather than on professional writers and peer-reviewed material, which makes Wikipedia similar to a social media platform. Unlike other
encyclopedias, Wikipedia is unrestricted in size and range of topics covered, and thereby holds
the potential to become the most comprehensive repository of human knowledge. Although
many studies have raised concerns about reliability and accuracy of Wikipedia content \cite{29,30,31}
for many people, not excluding scientists, Wikipedia is the first port of call for quick superficial
information search: 29.6\% of academics prefer Wikipedia to online library catalogues \cite{32}, and
52\% of students are frequent Wikipedia users, even if the instructor advised against it \cite{33}. In
general, browsing Wikipedia is the third most popular online activity, after watching YouTube
videos and engaging into social networking: it attracts 62\% of Internet users under 30 \cite{34}. The
popularity of Wikipedia seems to be facilitated by the Google search engine itself. A recent
study has found that in 96\% of cases Wikipedia ranks within the top 5 UK Google search
results \cite{35}, which means that Wikipedia content becomes (1) highly visible, taking a direct part
in shaping public opinion on a variety of topics, and (2) virtually unavoidable, whether the user
was searching for it or not.

In the present study, we wanted to investigate how the academia itself is represented on
Wikipedia. Previous research suggests that editing Wikipedia can be an influential way of
improving researchers' visibility or getting the message across, even in academic community \cite{4}.
Although there has not been sufficient research on exactly what it means if a scholar has a
Wikipedia page, it is considered prestigious to have one. According to an online survey
conducted by Nature, nearly 3\% of scholars have edited their Wikipedia biographies, and about
25\% check Wikipedia for references to themselves or their work \cite{4}. The decisions on the
inclusion/deletion of the articles in the encyclopedia are adopted through the consensus among
the editors, rather than imposed by a controlling institution. The articles need to satisfy some
notability criteria in order to be deemed worthy of inclusion, and in most cases are speedily
deleted if the community of editors deems them irrelevant \cite{36}. Previous research has
demonstrated that the topical coverage of Wikipedia is driven by the interests of its users, and
its comprehensiveness is likely to vary depending on the topic \cite{31}. Moreover, the cultural
preferences of the community of Wikipedia editors introduce additional subjectivity in the
editorial process of the encyclopedia \cite{37}.

A few studies have examined Wikipedia coverage of academically related topics. Elvebakk
compared Wikipedia coverage of 20th century philosophers with two peer-reviewed Web
encyclopedias and concluded that through the inclusion of “minor” and amateur philosophers,
Wikipedia gives a messier, more dynamic picture of the field, which, however, is not
fundamentally different from more traditional sources, but shows a slight tendency to a more
“popular” understanding of the discipline \cite{38}. Another qualitative evaluation of Wikipedia was
done by the experts in Communication studies who examined the encyclopaedia's articles on
communication research and revealed that Wikipedia is missing the contemporary research and
offers an incomplete and faulty impression on the current state of communication studies \cite{39}.
To the best of our knowledge, the only quantitative study that examined Wikipedia in academic context argues that among Computer
Science related topics and authors, those ones mentioned in the encyclopedia are more likely to have
higher academic and societal impact \cite{40}. Yet, we see some fallacies in these results
obtained by Jiang~et~al. Firstly, the reported Spearman correlation coefficient between
academic and Wikipedia ranking of authors is close to zero, which implies very low tendency for
one to predict the other. Secondly, their selection of authors is limited to those Computer
Scientists mentioned in the ACM Digital Library papers, which makes generalising the findings
to other fields and academia as a whole impossible. Lastly, the problem of name ambiguity was not addressed in the study.

Overall, the existing research is based on small samples limited to one discipline, and offers a
fragmented view of Wikipedia coverage of academic topics. This does not allow drawing
holistic conclusions about the role of the encyclopedia in both formal and informal academic
communication. To further investigate this matter, we examined 400 biographical Wikipedia
articles on living academics from the fields of (1) Biology, (2) Physics, (3) Computer Science,
(4) Psychology and Psychiatry. The articles differed in comprehensiveness and structure of
contents, but generally included researchers' short biography and sections covering their
personal and public life, research activities, scientific contributions, affiliation with institutions,
awards, etc. We tested the correlation between such parameters of the Wikipedia articles as
length, number of views, editors, edits, etc., and citation indexes of the academics, such as
total number of publications, total number of citations, $h$-index, etc. retrieved from the bibliographical database Scopus. The
analysis of the data allowed us to identify whether the researchers featured on Wikipedia have
high academic notability in their fields. To complete the picture, we also examined the
Wikipedia coverage of scientists introduced as “influential” by Thomson Reuters, a world
leading expert in bibliometrics and citation analyses.

\section*{Results}
Out of 400 randomly selected English Wikipedia articles on researchers (see Methods), 91\% of scholars
were academically active and had Scopus profiles (87 Biologists, 94 Computer Scientists, 98
Physicists, and 86 Psychologists and Psychiatrists). Of the remaining 9\% with no Scopus
record of publications, all had some relevant academic experience; namely, 34\% changed
occupation after completing their degrees; 31\% contribute to popularising science in their fields;
29\% are active academics, and 6\% have retired. The field-specific average Scopus metrics are
summarised in Table~\ref{t1}. On average, biologists in the sample are the most prolific authors,
accumulating the highest number of citations per document, total citations, and $h$-indexes. Computer Scientists have
the fewest average citations per document and demonstrate the lowest $h$-indexes in the sample.
Psychologists and Psychiatrists scholars on average have longest careers and collaborate with fewer co-authors than
researchers from other fields, generally producing the smallest number of documents.

\begin{figure}[h!]\includegraphics[width = .9 \linewidth]{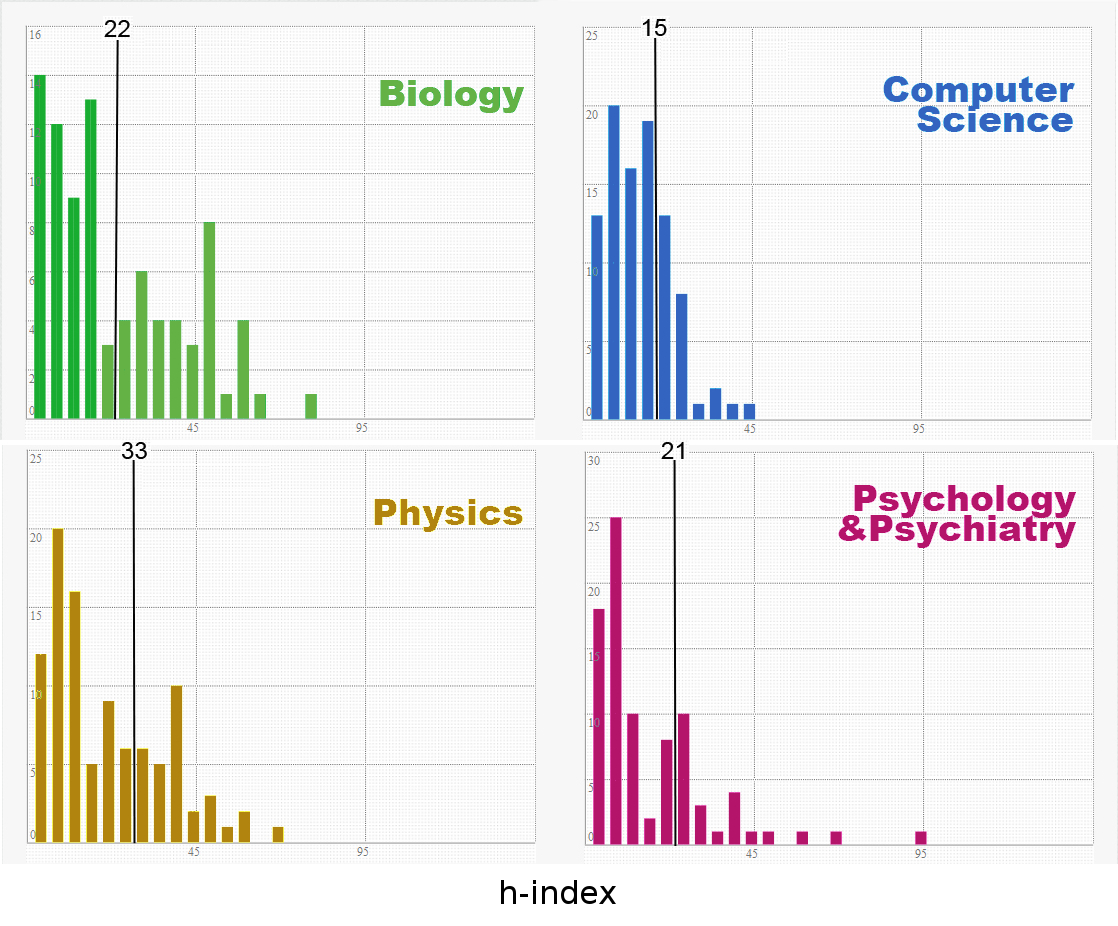}
  \caption{\csentence{Histograms of the $h$-indexes of the authors in the Wikipedia samples.}
      The black line indicates the average $h$-index in the field taken from the previous research \cite{11,41}.}\label{f1}
      \end{figure}

We compared the $h$-indexes of researchers from the Wikipedia samples who have Scopus profiles with the overall average $h$-indexes
by field established in previous research \cite{11,41}. The researchers, whose $h$-index was higher than the
field average, were considered notable. Figure~\ref{f1} demonstrates the histograms of $h$-indexes in the
observed fields and their distribution in relation to the field-specific average $h$-indexes. The analysis
has shown that only a small percentage of researchers mentioned on Wikipedia (36\% of Biologists,
31\% of Computer Scientists, 24\% Psychologists and Psychiatrists, and 22\% Physicists) are notable according to the traditional means of
evaluation (citation indexes). Table~\ref{t2} summarises the averages of Wikipedia articles metrics, and reveals field-specific differences
in the appearance and popularity of the articles. The articles on Physicists have the most diverse
coverage in other language editions of Wikipedia, are the longest in the sample, and on average
attract the largest number of unique editors and edits, which suggests a heightened interest of
Wikipedians in covering this topic. The articles on Computer Scientists, despite being the shortest, have the highest
in-degree and thus, are most connected with other Wikipedia articles. The articles on Psychologists and Psychiatrists are on
average the most viewed ones. 

\begin{figure}[h!]\includegraphics[width = .9 \linewidth]{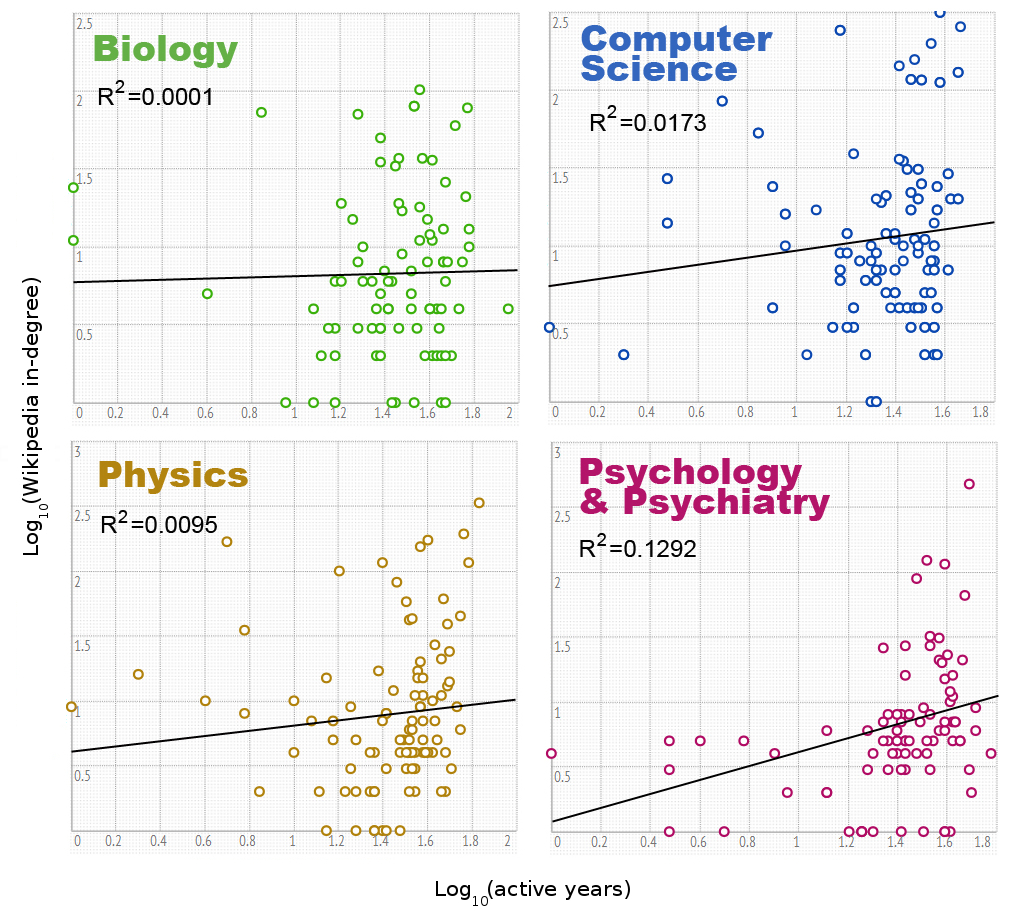}
  \caption{\csentence{ Scatterplots of the Wikipedia in-degree of the articles vs. the number of active years.}
      The solid line shows the best linear fit (using the least squares algorithm) to the data in log-log scale. The
strongest positive correlation is found in the Psychologists and Psychiatrists subset ($R^2$=0.1292).}\label{f2}
      \end{figure}

For all scholars with Scopus IDs, we examined 6$\times$8 binary
permutations of Scopus and Wikipedia metrics (presented in Tables~\ref{t1}~and~\ref{t2}), and performed regression analysis
in logarithmic space. All the calculated correlation coefficients are presented in Additional Table A1. Although we tried to consider all the potentially correlated pairs of parameters, to make sure that 
we do not miss any aspect of relation, but we discovered no significant correlation between any of the pairs. The strongest positive correlation ($R^2=0.13$) in the
dataset was found between in-degree (Wikipedia) and years active (Scopus) pair of variables in Psychologists and Psychiatrists
subset (shown in Figure~\ref{f2}). The strongest positive correlations in other subsets included the following pairs: number of language editions
(Wikipedia) vs. years active (Scopus) in Biologists ($R^2=0.019$), in-degree (Wikipedia) vs. mean
citations (Scopus) in Computer Scientists ($R^2=$0.023), and number of languages editions (Wikipedia) vs. mean citations (Scopus) in
Physics ($R^2=$0.035).

To investigate the coverage of the prominent researchers by Wikipedia, we also analysed a random
sample of 219 academics selected from the Thomson Reuters list of 1317 people “behind the world's
most influential research” \cite{42}. The results demonstrated that Wikipedia left out 52\% of the most
prominent Computer Scientists, 62\% of the most influential Psychologists and Psychiatrists, as well as 67\% of Biologists and Biochemists. The
poorest coverage was observed in Physics: 78\% were not found on Wikipedia. Importantly, the list
of highly cited researchers was published in 2010, whereas our data describe the current state of
Wikipedia, having left sufficient time to Wikipedia editors for a proper inclusion of the listed
prominent scientists. Despite this, Wikipedia has no record of the majority of the examined
scientists regardless of their field.

In order to measure the overall performance of Wikipedia in representing the notable scholars, we
calculated the $F$-scores for each of the examined fields (Table~\ref{t3}).
$F$-score is a measure of a classification accuracy based on the harmonic mean of precision (the proportion
of retrieved instances that are relevant, in this case the ratio of researchers on Wikipedia with $h$-index above field average)
and recall (the proportion of relevant instances that are
retrieved, in this case the percentage
of Highly Cited  researchers covered  by Wikipedia); $F=2 \times \frac{ \rm{precision} \times \rm{recall}}{\rm{precision}+\rm{recall}}$. 

$F$-score = 1 shows a perfect match between the model and the data, and a null model of
random assignment leads to $F=0.5$. Table~\ref{t3} shows that in each of the four inspected fields, the F-scores were
below the accuracy of a random assignment. In other
words, Wikipedia failed in both of the examined dimensions: being on Wikipedia does not signify
academic notability, and being notable does not guarantee Wikipedia coverage. Surprisingly, smaller
categories sometimes demonstrated higher recall than the bigger categories, which indicates that the 
growth of categories in terms of the number of articles, does not necessarily lead to a more
inclusive coverage.

      \begin{table}[h!]
\caption{Average citation metrics of $4 \times 100$ researchers randomly sampled from the
corresponding Wikipedia categories (bibliographical data are taken from Scopus). The last
column shows the proportion of researchers in the sample whose $h$-index was above the field
average (field-specific averages of $h$-indexes were taken from previous research \cite{11,41}). The numbers
in parentheses are the standard errors of the average.}
      \begin{tabular}{p{.5cm}p{1.3cm}p{1.7cm}p{1.2cm}p{1.cm}p{1.3cm}p{1.2cm}|p{1cm}||p{1cm}}
        \hline
Field & \# Papers& Citations & Citation \linebreak per \linebreak Paper& $h$-index & \# \linebreak co-authors & Years \linebreak active & Field ave. \linebreak $h$-index & $h$-index above field ave.  \\ \hline
Biol & 166 ($\pm$24) & 5,612 ($\pm$932) & 33 ($\pm$4) & 23 ($\pm$2) & 92 ($\pm$6) & 32 ($\pm$2) & 22 & 36\% \\
CoSi & 92 ($\pm$9) & 1,638 ($\pm$219) & 19 ($\pm$2) & 13 ($\pm$1) & 75 ($\pm$5) & 25 ($\pm$1) & 15 & 31\% \\
Phys & 139 ($\pm$16) & 3,512 ($\pm$447) & 27 ($\pm$4) & 19 ($\pm$2) & 95 ($\pm$6) & 31 ($\pm$1) & 33 & 22\% \\
Psyc  & 78 ($\pm$14) & 3,081 ($\pm$623) & 30 ($\pm$4) & 15 ($\pm$2) & 60 ($\pm$6) & 40 ($\pm$11) & 21 & 24\% \\ \hline
      \end{tabular}\label{t1}
\end{table}

\begin{table}[h!]
\caption{Average metrics of Wikipedia articles on researchers from the selected fields.
The length of a page is taken as 2,638 characters (a standard A4 Word document, Arial 12, 1.5
spaced). Languages column includes the number of language editions of Wikipedia covering the
same scholar (excluding English). The numbers in parentheses are the standard errors of the average.}
      \begin{tabular}{p{.5cm}p{1.3cm}p{3cm}p{3cm}p{1.2cm}p{1.3cm}p{1.2cm}}
        \hline
         Field&Length (page)&Editors/excluding bots&Edits/excluding bots&In-degree&Daily views(2012)&Languages\\ \hline
Biol&2.2($\pm$0.2)&27/22 ($\pm$3/$\pm$3)&54/47 ($\pm$6/$\pm$6)&15 ($\pm$2)&11 ($\pm$2)&2 ($\pm$0.4)\\
CoSi&2.0 ($\pm$0.1)&34/28 ($\pm$4/$\pm$4)&59/51 ($\pm$6/$\pm$6)&30 ($\pm$6)&19 ($\pm$4)&2 ($\pm$0.5)\\
Phys&2.7 ($\pm$0.3)&37/30 ($\pm$5/$\pm$4)&87/77 ($\pm$20/$\pm$19)&24 ($\pm$5)&17 ($\pm$3)&3 ($\pm$0.7)\\
Psyc&2.5 ($\pm$0.2)&34/29 ($\pm$6/$\pm$6)&76/69 ($\pm$14/$\pm$14)&16 ($\pm$5)&23 ($\pm$6)&1 ($\pm$0.3)\\ \hline

      \end{tabular}\label{t2}
\end{table}

\begin{table}[h!]
\caption{Overview of Wikipedia's accuracy in representing notable scholars.
{\it Precision} is the ratio of researchers on Wikipedia with $h$-index above field average and {\it recall} is the percentage
of Highly Cited  researchers covered  by Wikipedia. $F$-score is the harmonic mean of these two.}
      \begin{tabular}{p{2.5cm}p{1.2cm}p{1cm}p{1cm}|p{1.9cm}}
        \hline
        Wikipedia category& Precision&  Recall&$F$-score&Total number \linebreak of articles in\linebreak category\\ \hline
Biologists&36\%&33\%&0.34&81631\\
Computer Scientists&31\%&48\%&0.37&13789\\ 
Physicists&22\%&22\%&0.22&4554\\
Psychologists and\linebreak Psychiatrists&24\%&38\%&0.29&4777\\ \hline
      \end{tabular}\label{t3}
\end{table}

\section*{Discussion}
Wikipedia notability guidelines for academics suggest that the encyclopedia should only cover the
researchers who have “made significant impact in their disciplines”, which “in most cases” is
associated with being “an author of highly cited academic work” \cite{43}. Consequently, in the eyes of a lay
user, the mere presence of a scientist on Wikipedia is associated with their academic prestige and
authority. However, we observed
that, contrary to the expectations, the majority of the researchers' biographies on Wikipedia do not
meet the primary notability criteria and thus, their inclusion in the encyclopedia can be deceiving. Previous
qualitative examination of Wikipedia articles on certain academics \cite{38} and fields \cite{39} suggested that the 
selection of authors and topics by academic community differs from the one suggested by the
community of Wikipedians. Our findings show that this claim can be extended to a wider scope of fields
and authors, establishing that Wikipedia offers a very different image of researchers on the front
end of the scientific progress.

Our findings suggest that the inspection of Wikipedia is not useful in finding highly cited researchers.
Moreover, counter intuitively we observed that in some cases, the articles on authors without a Scopus track of scientific
publications were longer and attracted more editors than the articles on the scientists with high
citation indexes. This implies that the decisions of WP editors about covering certain scholars were motivated 
by the reasons other than the prominence of those scholars' bibliographic records.
Moreover, Wikipedia metrics of the articles about the
prominent researchers (with high $h$-indexes) were not statistically larger than the same metrics
from the less cited subset (Welch's t-test did not reject the null hypothesis of identical distributions;
the smallest p-value among all metrics and disciplines was 0.18). Consequently, we establish that
for a non-professional reader who turns to Wikipedia with an exploratory purpose of finding some
prominent researchers in a field, the encyclopedia might be misleading, as it provides no reliable
visual cues that might be a proxy of academic notability. We conclude
that the absence of correlation between Scopus and Wikipedia metrics suggests that they measure
different phenomena. As such, unlike other social media like Twitter and Facebook \cite{44}, Wikipedia
cannot be used as an early indicator of academic impact. That comes as a surprise, especially
considering that previous research has shown that Wikipedia activity data is a better predictor of
financial success of movies than Twitter \cite{45}. Yet, openness, speed, diversity, and collaborative filtering
offered by Wikipedia, can be applied to measure other aspects of scientific impact that are not
captured by the traditional citation analysis, for example social impact \cite{40} or public engagement of a
scientist.

We also investigated which proportion of truly prominent scientists (according to the ISI Highly Cited
Research list of most notable scientists) have Wikipedia presence, and discovered that the coverage
is below 50\% in each of the four examined fields. Since the list has been publically available since
2010, this observation cannot be due to time constraints. We know from previous research that
Wikipedia topical coverage is uneven and driven purely by the interests of its editors community \cite{31}.

Our findings establish that academic prominence of researchers (measured by citation counts) is not
among the factors facilitating the decisions on articles inclusion. Instead, the interest of Wikipedians
might be driven by other factors like scientists' social impact, public outreach, attention from media,
popularity of their research topic, etc. Interestingly, we observe that the academics with very low 
$h$-indexes and high Wikipedia visibility (measured in gained views) are all noted figures, book authors, and popularisers of science. 
This democratising effect of Wikipedia and Web 2.0 gives
young and promising academics more chance to be seen and found. On the other hand, it introduces
a subjective element into the hitherto structured and well-established system of peer-review-based
academic evaluation. Despite Wikipedia's inconsistency with the traditional view of scientific impact,
its content is highly visible and virtually unavoidable. The encyclopedia is making its way into
society, playing a role in forming public image on a variety of issues, not excluding science; and this
rise of Wikipedia is difficult to ignore. As the articles are being actively edited and viewed, individual
scientists, their fields, and entire academic institutions, can be easily affected by the way they are
represented in this important online medium.

One of the limitations of the present study relates to the inherent characteristics of bibliographical
database Scopus. Its citation metrics are restricted to the pool of 12,850 reviewed journals and do
not cover any publications before 1966 \cite{46}. This study only scrutinises one aspect of scholarly
notability – the citation metrics. Future studies might focus on testing whether other aspects – for
example, prestigious academic awards, membership in highly selective scholarly societies, the
impact of the work in the area of higher education and outside academia – raise the likelihood of
being included into Wikipedia. It could be instructive to qualitatively study the talk pages of
Wikipedia articles on academics in order to understand the motives for the inclusion/deletion of the
articles, and examine how the editors perceive the notability of the scientists covered. Future work
can also scrutinise if the presence on Wikipedia serves as a proxy to academics' social impact or
public visibility. And more importantly: Who is writing Wikipedia articles on academics (lay users,
academics, or the subjects of the article and their immediate social surroundings)? What motivates
their selection choices? How do the readers perceive the articles on academics? More research is
also needed to understand how the online image of scientists affects their reputation offline, as well
as the decisions regarding funding, conference invitations, grant allocation, collaboration, and
promotion. And specifically, what role does Wikipedia play in shaping this online image?

\section*{Methods}
\subsection*{Data Sources and Collection}
\subsubsection*{Wikipedia}
We used English Wikipedia's internal category tree structure to retrieve the full lists of
articles in the following categories: Biologists (81,631 articles), Physicists (4,554 articles), Computer
Scientists (13,789 articles), Psychologists and Psychiatrists (4,777 articles). From each of the lists,
we took a random sample of 300 entries and manually selected the first 100 articles that met the
following criteria: (1) the researcher was alive at the moment of data collection; (2) the article page
had no warning of a problem with Wikipedia notability guidelines for biographies; (3) the content of
the article and the assigned category suggested the same field affiliation. This left us with a dataset
of 400 articles. The article metrics were collected using SQL access to Wikimedia Toolserver
database and included: page ID, number of unique editors and edits, number of editors and edits
excluding those made by bots (``bot'' is a piece of code that runs through Wikipedia to implement
minor edits and other repetitive operations that help maintain the quality of Wikipedia articles \cite{47}),
length, in-degree (in this case, in-degree refers to the number of Wikipedia articles linking to the
selected article), number of page views, and number of other language editions of Wikipedia
in which the article was covered. 
\subsubsection*{Scopus}
We searched the academics from the 400 selected Wikipedia articles in Scopus
bibliographical database, and in cases where they were available, collected the following citation
metrics: author ID; number of documents, citations, co-authors, and years active; $h$-index, and
mean citations per paper. The data were collected and verified manually in order to exclude the
name ambiguity problem. In cases when the same researcher had two profiles, the citation data
were taken from the profile with most citations.

The complete datasets are available online in the Additional Dataset A1.
\subsubsection*{Prominent researchers}
To identify the researchers behind the most fundamental contributions to
the advancement of science, we used the data from the ISI Highly Cited Research study by
Thomson Reuters \cite{42}. The study is based on the top cited publications covered in Web of Science from
1981-2008, and is freely available online. We downloaded the list of the prominent researchers in
each of the four relevant fields (sampling frames) from http://highlycited.com/ website. The entries
were arranged alphabetically and consisted of researcher's name, surname, and organization. From
each of the sampling frames, we extracted every sixth entry and obtained four systematic random
samples: Biology and Biochemistry (49 researchers); Computer Science (63 researchers); Physics
(54 researchers); Psychology and Psychiatry (53 researchers). Then each author was searched in
Wikipedia to check whether there was a corresponding article. All data were collected in August
2013. The list of ``prominent researchers'' is available online in the Additional Dataset A2.
\subsection*{Data Analysis}
The data were imported into MATLAB to test the possible correlations between Wikipedia and Scopus
statistics. Histograms of all variables were visually inspected for normality and the data were
logarithmically transformed to compensate for the skewness of the data distribution. We built a
linear regression model and calculated the coefficient of determination $R^2$ to measure the strength of
association between all possible permutations of variables. The researchers with no Scopus IDs
were examined as separate cases.

%%%%%%%%%%%%%%%%%%%%%%%%%%%%%%%%%%%%%%%%%%%%%%
%%                                          %%
%% Backmatter begins here                   %%
%%                                          %%
%%%%%%%%%%%%%%%%%%%%%%%%%%%%%%%%%%%%%%%%%%%%%%

\begin{backmatter}

\section*{Competing interests}
  The authors declare that they have no competing interests.

\section*{Author's contributions}
AS and TY conceived and designed the research; AS analysed the data and wrote the main manuscript text. All authors
reviewed the manuscript.
\section*{Acknowledgements}
We would like to thank Ralph Schroeder, Eric T. Meyer (University of Oxford), Jasleen Kaur, and Filippo Radicchi (Indiana University) for insightful discussions. 
We also thank Wikimedia Deutchland~e.V. and Wikimedia Foundation for 
the live access to the Wikipedia data via Toolserver.

\bibliographystyle{bmc-mathphys} % Style BST file
\bibliography{biblio}      % Bibliography file (usually '*.bib' )

%%%%%%%%%%%%%%%%%%%%%%%%%%%%%%%%%%%
%%                               %%
%% Additional Files              %%
%%                               %%
%%%%%%%%%%%%%%%%%%%%%%%%%%%%%%%%%%%
\newpage

\section*{Additional Files}
%  \subsection*{Additional file 1 (.pdf) --- Table A1}

 \begin{figure}[h!]\includegraphics[width = .97 \linewidth]{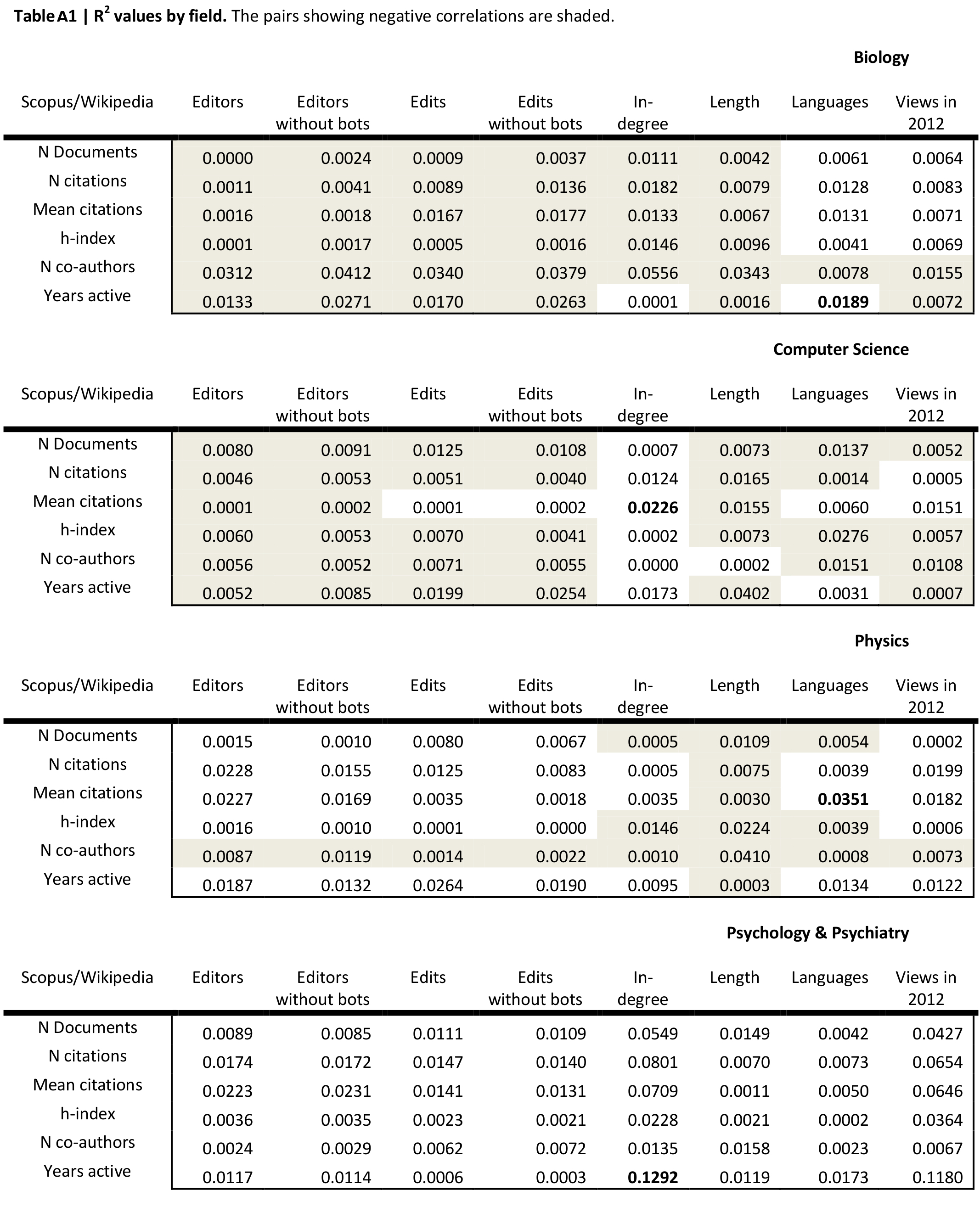}
 %\caption{\csentence{ }\label{fA1}
      \end{figure}

  \subsection*{Additional file 2 (.xls) --- Dataset A1}
    The dataset A1 contains the complete list of Wikipedia metrics and Scopus data for all four
disciplines.

  \subsection*{Additional file 3 (.xls) --- Dataset A2}
    The Dataset A2 contains the random sample of prominent researchers by field taken from the “ISI
Highly Cited Research” by Thomson Reuters.

\end{backmatter}
\end{document}